\documentstyle[ltwol,epsf]{article}

\def\beq{\begin{equation}}
\def\eeq{\end{equation}}
\def\bea{\begin{eqnarray}}
\def\eea{\end{eqnarray}}
\def\nn{\nonumber}
\def\sss{\scriptscriptstyle}

\def\bd{B_d^0}
\def\bdbar{{\overline{B_d^0}}}
\def\bs{B_s^0}
\def\bsbar{{\overline{B_s^0}}}
\def\bbar{{\overline{B^0}}}
\def\barp{{\raise.35ex\hbox
{${\sss (}$}}---{\raise.35ex\hbox{${\sss )}$}}}
\def\bdbarp{\hbox{$B_d$\kern-1.4em\raise1.4ex\hbox{\barp}}}
\def\bsbarp{\hbox{$B_s$\kern-1.4em\raise1.4ex\hbox{\barp}}}
\def\barpd{{\raise.35ex\hbox
{${\sss (}$}}--{\raise.35ex\hbox{${\sss )}$}}}
\def\dbarp{\hbox{$D^{*0}$\kern-1.6em\raise1.5ex\hbox{\barpd}}}
\def\kbarp{\hbox{$K^{*0}$\kern-1.6em\raise1.5ex\hbox{\barpd}}}
\def\ks{K_{\sss S}}

\def\roughly#1{\mathrel{\raise.3ex\hbox
{$#1$\kern-.75em\lower1ex\hbox{$\sim$}}}}

\def\adir00{{a_{\sss dir}^{00}}}
\def\B00{B^{00}}
\def\Bp0{B^{+0}}

\def\epjc#1#2#3{{\it Eur.\ Phys.\ J.}\ {\bf C#1}, #3 (19#2)}
\def\epjcn#1#2#3{{\it Eur.\ Phys.\ J.}\ {\bf C#1}, #3 (20#2)}

\def\npb#1#2#3{{\it Nucl.\ Phys.} {\bf B#1}, #3 (19#2)}
\def\plb#1#2#3{{\it Phys.\ Lett.} {\bf #1B}, #3 (19#2)}
\def\prd#1#2#3{{\it Phys.\ Rev.} {\bf D#1}, #3 (19#2)}

\def\newprdtwo#1#2#3{{\it Phys.\ Rev.} {\bf D#1}: #3 (20#2)}

\def\prl#1#2#3{{\it Phys.\ Rev.\ Lett.} {\bf #1}, #3 (19#2)}
\def\zpc#1#2#3{{\it Zeit.\ Phys.} {\bf C#1}, #3 (19#2)}
\def\ijmp#1#2#3{{\it Int.\ J.\ Mod.\ Phys.} {\bf A#1}, #3 (19#2)}

\begin{document}

\title{Extracting Weak Phase Information from $B \to V_1 V_2$ Decays}

\author{David London} 

\address { Laboratoire Ren\'e J.-A. L\'evesque, 
Universit\'e de Montr\'eal,\\ 
C.P. 6128, succ.\ centre-ville, Montr\'eal, QC, Canada H3C 3J7\\
E-mail: london@lps.umontreal.ca}

\author{Nita Sinha and Rahul Sinha}

\address {The Institute of Mathematical Sciences,
 Taramani, Chennai 600 113, India\\
E-mail: nita@imsc.ernet.in, sinha@imsc.ernet.in}  


\twocolumn[\begin{flushright}
\vskip -1.cm
UdeM-GPP-TH-01-85\\IMSc/2001/04/17\\
\end{flushright}
\maketitle\abstracts{ We describe a new method for
  extracting weak, CP-violating phase information, with no hadronic
  uncertainties, from an angular analysis of $B \to V_1 V_2$ decays,
  where $V_1$ and $V_2$ are vector mesons. The quantity $\sin^2
  (2\phi_1 + \phi_3)$ can be obtained cleanly from the study of decays
  such as $\bd(t) \to D^{*\pm} \rho^\mp$, $D^{*\pm} a_1^{\mp}$,
  $\dbarp~~\kbarp ~~$, etc.  Similarly, one can use $\bs(t) \to
  D_s^{*\pm} K^{*\mp}$ or even $B^\pm\to\dbarp ~~K^{*\pm}$ to extract
  $\sin^2 \phi_3$.  There are no penguin contributions to these
  decays. It is possible that $\sin^2 (2\phi_1 + \phi_3)$ will be the
  second function of CP phases, after $\sin 2\phi_1$, to be measured
  at $B$-factories.  }]

One of the most important open questions in particle physics is the
origin of CP violation. In the standard model (SM), CP violation is
due to the presence of a nonzero complex phase in the
Kobayashi-Maskawa (KM) quark mixing matrix. This explanation can be
tested in the $B$ system by measuring the CP-violating rate
asymmetries in $B$ decays, thereby extracting $\phi_1(\equiv \beta)$, $\phi_2(\equiv \alpha)$ and
$\phi_3(\equiv \gamma)$, the three interior angles of the unitarity
triangle.~\cite{CPreview}

The reason that $B$ decays are such a useful tool is that the CP
angles can be obtained without hadronic uncertainties. 
%
%
It was thought that the CP angles could be easily measured in $\bd (t)
\to J/\psi\ks ~(\phi_1)$, $\bd(t) \to \pi^+ \pi^- ~(\phi_2)$, and
$\bs(t) \to \rho\ks ~(\phi_3)$.  It has become clear that the presence
of penguin amplitudes~\cite{penguins} makes the extraction of $\phi_2$
from $\bd(t) \to \pi^+ \pi^-$ more difficult, and completely spoils
the measurement of $\phi_3$ in $\bs(t) \to \rho\ks$. Even in the
gold-plated mode $\bd(t) \to \Psi\ks$, penguin contributions limit the
precision with which $\phi_1$ can be measured to about 2\%. A great
deal of work has since been done developing new methods to cleanly
obtain the CP angles from a wide variety of final states.

One class of final states that was considered consists of two vector
mesons, $V_1 V_2$. Because the final state does not have a
well-defined orbital angular momentum, the final state $V_1 V_2$
cannot be a CP eigenstate. This then implies that, even if both $B^0$
and $\bbar$ can decay to the final state $V_1 V_2$, one cannot extract
a CP phase cleanly. However, this situation can be remedied with the
help of an angular analysis.~\cite{helicity} By examining the decay
products of $V_1$ and $V_2$, one can measure the various helicity
components of the final state. Since each helicity state corresponds
to a state of well-defined CP, an angular analysis allows one to use
$B \to V_1 V_2$ decays to obtain one of the CP phases cleanly.

We show that the angular analysis is more powerful than has been
realized previously. Due to the interference between the different
helicity states, there are enough independent measurements that one
can obtain weak phase information from the decays of $B^0$ and $\bbar$
to a common final state $f$. Furthermore, contrary to other methods,
it is not necessary to measure the branching ratios of both $B^0 \to
f$ and $\bbar \to f$. This is important for final states such as
$D^{*\pm} \rho^{\mp}$, in which one of the two decay amplitudes is
considerably smaller than the other one.

\begin{figure}[tb]
\vspace{-0.28in}
\centerline{\epsfxsize 3.0truein \epsfbox {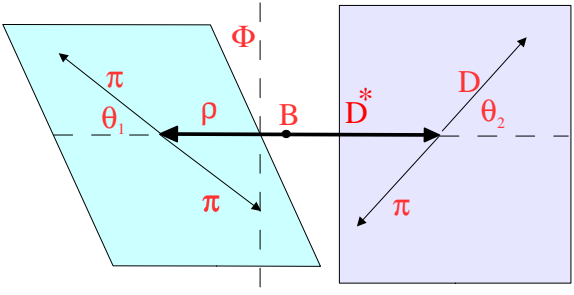}}
\caption{The angular distribution of the decay $B\to D^*\rho\to
(D\pi)(\pi\pi)$}
\label{fig:ang-dist}
\end{figure}

The most general covariant amplitude for a $B$ meson decaying to a
pair of vector mesons has the form
\begin{eqnarray} 
&& A(
\displaystyle
 B(p)\to V_1(k,\epsilon_1) V_2(q,\epsilon_2))= 
\displaystyle
\epsilon_1^{*\mu}\epsilon_2^{*\nu} \times \nn \\[2ex] &&
\left(
 a\; g_{\mu\nu}+ 
\frac{b}{m_1 m_2} p_{\mu} p_{\nu} 
+i\frac{c}{m_1 m_2} 
\epsilon_{\mu\nu\alpha\beta} k^\alpha q^\beta 
\right)~,
\end{eqnarray} 
where $m_1$, $m_2$ are the masses of $V_1$, $V_2$ respectively. The
coefficients $a$, $b$, and $c$ can be expressed in terms of the linear
polarization basis $A_{\|}$, $A_\perp$ and $A_0$ as follows:
\begin{eqnarray}
A_0&=&- x a-(x^2-1) b ~, \nn \\
A_\|&=&\sqrt{2}a ~, \\
A_\perp&=&\sqrt{2(x^2-1)}\,c ~, \nn
\end{eqnarray}
where $x=k.q/(m_1 m_2)$.  If both mesons subsequently decay into two
$J^{P}\!=\!0^{-}$ mesons, {\it i.e.\/} $V_1\!\to\! P_1 P_1^\prime$ and
$V_2 \to P_2 P_2^\prime$, the amplitude can be expressed
as~\cite{flambda,flamdun}
\begin{eqnarray} 
&&A(\displaystyle B\to V_1 V_2)\propto (A^0 
\cos\theta_1 \cos\theta_2 + \frac{A^\|}{\sqrt{2}} \nonumber \\
&&\sin\theta_1 \sin\theta_2 \cos\phi 
-i \frac{A^\perp}{\sqrt{2} }\sin\theta_1
\sin\theta_2 \sin\phi), 
\label{ang-dist}
\end{eqnarray}
where $\theta_1$ ($\theta_2$) is the angle between the $P_1$ ($P_2$)
three-momentum vector, $\vec{k_1}(\vec{q_1})$ in the $V_1 ( V_2)$ rest
frame and the direction of total $V_1$ ($V_2$) three-momentum vector
defined in the $B$ rest frame.  $\Phi$ is the angle between the
normals to the planes defined by $P_1 P_1^\prime$ and $P_2
P_2^\prime$, in the $B$ rest frame. The angular distribution for the
decay $B\to D^*\rho$ is shown in Figure~\ref{fig:ang-dist}.

We consider a final state $f$, consisting of two vector mesons, to
which both $B^0$ and $\bbar$ can decay. If only
one weak amplitude contributes to $B^0 \to f$ and $\bbar \to f$, we
can write the helicity amplitudes as follows:
\begin{eqnarray}
\label{amp1}
A_\lambda \equiv Amp (B^0 \to f)_\lambda &=& 
       a_\lambda e^{i\delta_\lambda^a} e^{i\phi_a} ~, \\ 
A'_\lambda \equiv Amp (\bbar \to f)_\lambda &=& 
       b_\lambda e^{i\delta_\lambda^b} e^{i\phi_b} ~, \\ 
{\bar A}'_\lambda \equiv Amp (B^0 \to {\bar f})_\lambda &=& 
       b_\lambda e^{i\delta_\lambda^b} e^{-i\phi_b} ~, \\ 
{\bar A}_\lambda \equiv Amp (\bbar \to {\bar f})_\lambda &=& 
       a_\lambda e^{i\delta_\lambda^a} e^{-i\phi_a} ~,
\label{amp4}
\end{eqnarray}
where the helicity index $\lambda$ takes the values $\left\{
  0,\|,\perp \right\}$. In the above, $\phi_{a,b}$ and
$\delta^{a,b}_\lambda$ are the weak and strong phases, respectively.

Using CPT invariance, and the decay amplitude expressed in
Eq.~\ref{ang-dist}, the total decay amplitudes can be written as
\begin{eqnarray}
{\cal A} = Amp (B^0\to f) = A_0 g_0 + A_\| g_\| + i \, A_\perp g_\perp~~ ,
\label{A}\\
{\bar{\cal A}} = Amp (\bbar\to {\bar f}) = 
     {\bar A}_0 g_0 + {\bar A}_\| g_\| - i \, {\bar A}_\perp g_\perp~~ ,
\label{Abar}\\
{\cal A}' = Amp (\bbar\to f) = A'_0 g_0 + A'_\| g_\| - i \, A'_\perp g_\perp~~ ,
\label{Aprime}\\
{\bar{\cal A}}' = Amp (B^0 \to {\bar f}) = 
   {\bar A}'_0 g_0 + {\bar A}'_\| g_\| + i \, {\bar A}'_\perp g_\perp~~ ,
\label{Aprimebar}
\end{eqnarray}
where the $g_\lambda$ are the coefficients of the helicity amplitudes,
defined using Eq.~\ref{ang-dist}.

%
With the above equations, the time-dependent decay rate for a $B^0$
decaying into the two vector--meson final state, {\em i.e.\/} $B^0(t)
\to f$, can be expressed as
\bea
\Gamma(B^0(t) \to f)= e^{-\Gamma t} \sum_{\lambda\leq\sigma}&&
\Bigl(\Lambda_{\lambda\sigma} + \Sigma_{\lambda\sigma}\cos(\Delta M t)\nn\\&&
- \!\rho_{\lambda\sigma}\sin(\Delta M t)\Bigr) g_\lambda g_\sigma\!~.
\eea
By performing a time-dependent study and angular analysis of the decay
$B^0(t)\to\! f$, one can measure the 18 observables
$\Lambda_{\lambda\sigma}$, $\Sigma_{\lambda\sigma}$ and
$\rho_{\lambda\sigma}$. In terms of the helicity amplitudes
$A_0,A_\|,A_\perp$, these can be expressed as follows:
\bea
&\Lambda_{\lambda\lambda}=\displaystyle
\frac{|A_\lambda|^2+|A'_\lambda|^2}{2},&
\Sigma_{\lambda\lambda}=\displaystyle
\frac{|A_\lambda|^2-|A'_\lambda|^2}{2},\nn \\[1.5ex]
&\Lambda_{\perp i}= -\!{\rm Im}({ A}_\perp { A}_i^* \!-\! A'_\perp {A'_i}^* ),
&\Lambda_{\| 0}= {\rm Re}(A_\| A_0^*\! +\! A'_\| {A'_0}^* ),
\nn \\[1.5ex]
&\Sigma_{\perp i}= -\!{\rm Im}(A_\perp A_i^*\! +\! A'_\perp {A'_i}^* ),
&\Sigma_{\| 0}= {\rm Re}(A_\| A_0^*\!-\! A'_\| {A'_0}^* ),\nn\\[1.5ex]
&\rho_{\perp i}\!=\!-\!{\rm Re}\!\Bigl(\!\frac{q}{p}
\![A_\perp^*  A'_i\! +\! A_i^* A'_\perp\!]\!\Bigr),
&\rho_{{\sss \perp \perp}}\!=\! -\! {\rm Im}\Bigl(\frac{q}{p}\,
A_\perp^* A'_\perp\Bigr),\nn\\[1.5ex]
&\rho_{\| 0}\!=\!{\rm Im}\!\Bigl(\frac{q}{p}\!
[A_\|^* A'_0\! + \!A_0^* A'_\|\!]\!\Bigr),
&\rho_{ii}\!=\!{\rm Im}\!\Bigl(\frac{q}{p} A_i^* A'_i\Bigr),
\label{defs}
\eea
where $i=\{0,\|\}$. In the above, $q/p = \exp({-2\,i\phi_{\sss M}})$,
where $\phi_{\sss M}$ is the weak phase present in $B^0$--$\bbar$
mixing.

Similarly, the decay rate for $B^0(t) \to {\bar f}$ is given by
%
\bea
\Gamma(B^0(t)\to {\bar f}) = e^{-\Gamma t} \sum_{\lambda\leq\sigma}&&
\Bigl(
{\bar\Lambda}_{\lambda\sigma} + {\bar\Sigma}_{\lambda\sigma}\cos(\Delta M t) \nn\\&&
- {\bar\rho}_{\lambda\sigma}\sin(\Delta M t)
\Bigr)  g_\lambda  g_\sigma ~.
\eea
The expressions for the observables ${\bar\Lambda}_{\lambda\sigma}$,
${\bar\Sigma}_{\lambda\sigma}$ and ${\bar\rho}_{\lambda\sigma}$ are
similar to those given in Eq.~(\ref{defs}), with the replacements
$A_\lambda \to {\bar A}'_\lambda$ and $A'_\lambda \to {\bar
  A}_\lambda$.

With the above expressions for the various amplitudes, we now show how
to extract weak phase information using the above measurements. First,
we note that
\beq
\Lambda_{\lambda\lambda}={\bar\Lambda}_{\lambda\lambda}=
\frac{(a_\lambda^2+b_\lambda^2)}{2},
\Sigma_{\lambda\lambda}=-{\bar\Sigma}_{\lambda\lambda}=
\frac{(a_\lambda^2-b_\lambda^2)}{2}.
\label{LamSig_eq}
\eeq
Thus, one can determine the magnitudes of the amplitudes appearing in
Eqs.~(\ref{amp1})--(\ref{amp4}), $a_\lambda^2$ and $b_\lambda^2$.
However, it must be stressed that {\em the knowledge of $b_\lambda^2$
  will not be necessary within our method}. This is important for the
final states that have $b_\lambda \ll a_\lambda$, for which the
determination of $b_\lambda^2$ would be very difficult.

Next, we have
\bea
&&\Lambda_{\perp i}\! =\! -{\bar\Lambda}_{\perp i}\! =\! b_{\perp} b_i
\sin(\delta_{\perp}\!-\!\delta_i\!+\!\Delta_i)-a_\perp a_i\sin(\Delta_i), \nn \\
&&\Sigma_{\perp i}\! =\! {\bar\Sigma}_{\perp i}\! =\! -b_\perp b_i
\sin(\delta_\perp\!-\!\delta_i\!+\!\Delta_i)-a_\perp a_i\sin(\Delta_i),
\label{LS}
\eea
where $\Delta_i \equiv \delta_\perp^a-\delta_i^a$ and $\delta_\lambda
\equiv \delta_\lambda^b-\delta_\lambda^a$. Using Eq.~(\ref{LS}) one
can solve for $a_\perp a_i\sin\Delta_i$. We will see that this is the
only combination needed to cleanly extract weak phase information.

The coefficients of the $\sin(\Delta m t)$ term, which can be
obtained in a time-dependent study, can be written as
\beq
\rho_{\lambda\lambda}\! =\! \pm a_\lambda b_\lambda \sin(\phi\!+\!\delta_\lambda),
{\bar\rho}_{\lambda\lambda}\!=\!\pm a_\lambda b_\lambda \sin(\phi\!-\!\delta_\lambda),
\label{rho_eq}
\eeq
where the sign on the right hand side is positive for $\lambda=\|,0$,
and negative for $\lambda=\perp$. In the above, we have defined the CP
phase $\phi \equiv -2\phi_{\sss M} + \phi_b - \phi_a$. These
quantities can be used to determine
\beq
2 b_\lambda\cos\delta_\lambda\! =\! 
\pm\frac{\rho_{\lambda\lambda}\!+\!{\bar\rho}_{\lambda\lambda}}{a_\lambda \sin\phi},
2 b_\lambda\sin\delta_\lambda \!=\!
\pm\frac{\rho_{\lambda\lambda}\!-\!{\bar\rho}_{\lambda\lambda}}{a_\lambda \cos\phi}.
\label{delta}
\eeq
Similarly, the terms involving interference of different helicities
are given as
\begin{eqnarray}
\rho_{\perp i}\! &=&\! -a_\perp b_i \cos(\phi\!+\!\delta_i\!-\!\Delta_i)\!-\!
a_i b_\perp \cos(\phi\!+\!\delta_\perp\!+\!\Delta_i) , \nn\\
{\bar\rho}_{\perp i}\! &=&\! -a_\perp b_i  \cos(\phi\!-\!\delta_i\!+\!\Delta_i) \!-\! 
a_i b_\perp \cos(\phi\!-\!\delta_\perp\!-\!\Delta_i) .
\label{rhocombs}
\end{eqnarray}
%
%

Putting all the above information together, we are now in a position
to extract the weak phase $\phi$. Using Eq.~(\ref{delta}), the
expressions in Eq.~(\ref{rhocombs}) can be used to yield
\begin{eqnarray}
&&\!\rho_{\perp i}\!+\!{\bar\rho}_{\perp i}\!=
-\cot\phi\,{a_i a_\perp}\cos\Delta_i\Bigg[\frac{\rho_{i
i}+{\bar\rho}_{i i}}{a_i^2}- \frac{\rho_{\sss\perp
\perp}+{\bar\rho}_{{\sss \perp \perp}}}{a_\perp^2}\Bigg] \nn \\ &&
~~~~~~~~~~~~-{a_i a_\perp}\sin\Delta_i\Bigg[\frac{\rho_{i i}-{\bar\rho}_{i
i}}{a_i^2}+ \frac{\rho_{{\sss \perp \perp}}-{\bar\rho}_{
{\sss \perp \perp}}}{a_\perp^2}\Bigg],\\ 
&&\rho_{\perp i}\!-\!{\bar\rho}_{\perp i}=
\tan\phi\,{a_i a_\perp}\cos\Delta_i\Bigg[\frac{\rho_{i
i}-{\bar\rho}_{i i}}{a_i^2}- \frac{\rho_{\sss\perp
\perp}-{\bar\rho}_{{\sss \perp \perp}}}{a_\perp^2}\Bigg] \nn \\ &&
~~~~~~~~~~~~-{a_i a_\perp}\sin\Delta_i\Bigg[\frac{\rho_{i i}+{\bar\rho}_{i
i}}{a_i^2}+ \frac{\rho_{{\sss \perp \perp}}+{\bar\rho}_{
{\sss \perp \perp}}}{a_\perp^2}\Bigg].
\label{sol}
\end{eqnarray}
Now, we already know most of the quantities in the above two
equations: (i) $\rho_{\lambda\sigma}$ and ${\bar\rho}_{\lambda\sigma}$
are measured quantities, (ii) the $a_\lambda^2$ are determined from
the relations in Eq.~(\ref{LamSig_eq}), and (iii) ${a_i
  a_\perp}\sin\Delta_i$ is obtained from Eq.~(\ref{LS}). Thus, the
above two equations involve only two unknown quantities --- $\tan\phi$
and ${a_i a_\perp}\cos\Delta_i$ --- and can easily be solved (up to a
sign ambiguity in each of these quantities). In this way $\tan^2\phi$
(or, equivalently, $\sin^2 \phi$) can be cleanly obtained from the
angular analysis.

Note that our method relies on the measurement of the interference
terms between different helicities. However, we do not actually
require that all three helicity components of the amplitude be used.
In fact, one can use observables involving any two of largest helicity
amplitudes. In the above description, one could have chosen `$\|\,0$'
instead of `$\perp\!\|$' or `$\perp\! 0$'.

We now turn to specific applications of this method. Consider first
the situation in which the final state is a CP eigenstate, $f = \pm
{\bar f}$. In this case, the parameters of
Eqs.~(\ref{amp1})--(\ref{amp4}) satisfy $a_\lambda = b_\lambda$,
$\delta_\lambda^a = \delta_\lambda^b$ (which implies that
$\delta_\lambda = 0$), and $\phi_a = -\phi_b$ (so that $\phi \equiv
-2\phi_{\sss M} + 2 \phi_b$). As described above, $a_\lambda^2$ can be
obtained from Eq.~\ref{LamSig_eq}. But now the measurement of
$\rho_{\lambda\lambda}$ [Eq.~(\ref{rho_eq})] directly yields $\sin
\phi$. In fact, this is the conventional way of using the angular
analysis to measure the weak phases: each helicity state separately
gives clean CP-phase information. Thus, when $f$ is a CP eigenstate,
nothing is gained by including the interference terms.

Of course, in general, final states that are CP eigenstates will all
receive penguin contributions at some level.
Thus, these states violate our assumption that only one weak amplitude
contributes to $B^0 \to f$ and $\bbar \to f$. The only quark-level
decays which do not receive penguin contributions are ${\bar b} \to
{\bar c} u {\bar d} ,~ {\bar u} c {\bar d}$, as well as their
Cabibbo-suppressed counterparts, ${\bar b} \to {\bar c} u {\bar s} ,~
{\bar u} c {\bar s}$.
%
Consider first the decays $\bd/\bdbar \to D^{*-}\rho^+, D^{*+}\rho^-$
In this case we have $\phi_{\sss M} = \phi_1$, $\phi_a = 0$ and
$\phi_b = -\phi_3$, so that $\phi = - 2 \phi_1 - \phi_3$. The method
described above allows one to extract $\sin^2 (2\phi_1 + \phi_3)$ from
an angular analysis of the final state $D^{*\pm} \rho^\mp$.

In Ref.~6, Dunietz pointed out that $\sin^2 (2\phi_1 +
\phi_3)$ could, in principal, be obtained from measurements of $\bd(t)
\to D^\mp \pi^\pm$. He used the method of Ref.~7, which
requires the accurate measurement of the quantity $\Gamma(\bdbar \to
D^- \pi^+) / \Gamma(\bd \to D^- \pi^+)$. This ratio is essentially
$|V_{ub} V_{cd}^* / V_{cb}^* V_{ud}|^2 \simeq 4 \times 10^{-4}$.
Obviously, it will be very difficult to measure this tiny quantity
with any precision, which creates a serious barrier to carrying out
Dunietz's method in practice.  On the other hand, our method does not
suffer from this problem. In our notation
[Eqs.~(\ref{amp1})--(\ref{amp4})], the rate $\Gamma(\bdbar \to D^{*-}
\rho^+)$ is proportional to $b_\lambda^2$.  However, as we have
already emphasized in the discussion following Eq.~(\ref{LamSig_eq}),
a determination of this quantity is not needed to extract $\sin^2
(2\phi_1 + \phi_3)$ using the angular analysis: {\it none of the
  observables or combinations required for the analysis are
  proportional to $b_\lambda^2$}. Thus, we avoid the practical
problems present in Dunietz's method.

The two decay amplitudes for the final states $D^{*\pm} \rho^\mp$ have
very different sizes, {\it i.e.\/} $b_\lambda \ll a_\lambda$. This
results in a very small CP-violating asymmetry whose size is
approximately $|V_{ub} V_{cd}^* / V_{cb}^* V_{ud}| \approx 2\%$.
Thankfully, the situation is alleviated by the large branching ratio
for the decay $\bd \to D^{*-}\rho^+$, roughly 1\%.
The Cabibbo-suppressed decays, {\it e.g.\/} 
$\bd \!\to \!{\bar D}^{*0}\! K^{*0}$, $D^{*0} K^{*0}$ and
$\bdbar\!\to\! D^{*0} {\bar K}^{*0}$, ${\bar D}^{*0} {\bar K}^{*0}$,
with $K^{*0}$ and ${\bar K}^{*0}$ decaying to $\ks \pi^0$, lead to a
larger asymmetry of about $|V_{ub} V_{cs}^* / V_{cb}^* V_{us}| \approx
40\%$.~\cite{BDK} However, such Cabibbo-suppressed decays have much
smaller branching ratios than those for $\bd/\bdbar \to
D^{*\pm}\rho^\mp$.

One can also consider $\bs$ and $\bsbar$ decays.  corresponding to the
quark-level decays ${\bar b} \to {\bar c} u {\bar d} ,~ {\bar u} c
{\bar d}$, or ${\bar b} \to {\bar c} u {\bar s} ,~ {\bar u} c {\bar
  s}$.  The most promising processes are the Cabibbo-suppressed decay
modes $\bs/\bsbar \to D_s^{*\pm} K^{*\mp}$. Here the $\bs-\bsbar$
mixing phase is almost $0$, so that the quantity $\sin^2 \phi_3$ can
be extracted from the angular analysis of $\bs(t)\to D_s^{*\pm}
K^{*\mp}$. Other methods~\cite{ADK,DF} for obtaining the CP phase
$\phi_3$ using similar final states have also been
considered.~\cite{LSS2}

It is {\em even possible to cleanly extract the weak phase $\phi_3$
  using only charged $B^\pm$ decays,}~\cite{flambda} by studying the
angular distribution. The decays $B^+\to D^{*0} V^+$, $B^+\to
\overline{D^{*0}} V^+$ and $B^-\to D^{*0} V^-$, $B^-\to
\overline{D^{*0}} V^-$ can be related by CPT. Consider
$D^{*0}/\overline{D^{*0}}$ decaying into $D^0\pi^0 /
\overline{D^0}\pi^0$, with $D^0/\overline{D^0}$ meson further decaying
to a final state `$f$' that is common to both $D^0$ and
$\overline{D^0}$. $f$ is chosen to be a Cabibbo-allowed mode of $D^0$
or a doubly-suppressed mode of $\overline{D^0}$. The amplitudes for
the decays of $B^+$ and $B^-$ to a final state involving $f$ will be a
sum of the contributions from $D^{*0}$ and $\overline{D^{*0}}$, and
similarly for the CP-conjugate processes. In this case one can
experimentally measure the magnitudes of the 12 helicity amplitudes,
as well as the interference terms, leading to a total of 24
independent observables.  However, there are just 15 unknowns involved
in the amplitudes: $a_\lambda, b_\lambda, \delta_\lambda^a,
\delta_\lambda^b, \phi_3,\Delta~{\rm and}~{\cal R}$; where, ${\cal
  R}^2=\frac{Br(\overline {D^{0}}\to f)}{Br(D^{0}\to f)},$ and
$\Delta$ is the strong phase difference between $D^0\to f$ and $D^0\to
{\overline f}$. Hence, the weak phase $\phi_3$ may be cleanly
extracted.

The extraction of $\sin^2(2\phi_1+\phi_3)$ may well turn out to be the
second clean measurement to be made at $B$-factories. Studies are
already underway for a possible measurement at the first generation B
factory.~\cite{expts} The angle $\phi_2$ can be measured using an
isospin analysis in $\bd\to \pi\pi$,~\cite{isospin} but this technique
requires measuring the branching ratio for $\bd\to\pi^0\pi^0$, which
may be quite small. It is also possible to extract $\phi_2$ using a
Dalitz-plot analysis of $\bd(t)\to\pi^+\pi^-\pi^0$
decays.~\cite{Dalitz} It is estimated that this measurement will take
roughly six years to complete. As for the angle $\phi_3$, the original
suggestion using the decays $B^\pm \to D^0 K^\pm ,~ {\overline{D^0}}
K^\pm ,~ D^0_{\sss CP} K^\pm$~\cite{GroWyler} runs into problems
because it is virtually impossible to tag the flavor of the
final-state $D$-meson.
%
%
One can still obtain $\phi_3$ cleanly in other
modes~\cite{ADS,flambda} but this requires many more $B$'s, so that it
is unlikely such measurements can be carried out in the first
generation $B$-factories. Finally, there has been much work recently
looking at the possibilities for extracting $\phi_3$ from $B\to\pi K$
decays.~\cite{BpiKreview} However, all of these methods use flavor
$SU(3)$ symmetry, and so rely heavily on theoretical input. In view of
all of this, it is conceivable that the second clean extraction of CP
phases at $B$ factories will be the measurement of
$\sin^2(2\phi_1+\phi_3)$ using the method described here. We also note
that the measurement of $\sin^2(2\phi_1+\phi_3)$ may turn out to be
very useful in looking for physics beyond the SM. For more details one
is referred to Refs.~11 and 18.

To summarize, we have presented a new method of using an angular
analysis in $B\to VV$ decay modes, which do not receive penguin
contributions, to {\bf{\em cleanly extract}} the weak phases $(2
\phi_1+\phi_3)$ and $\phi_3$.  We have shown that the quantity $\sin^2
(2\phi_1 + \phi_3)$ can be cleanly obtained from the time dependent
angular analysis study of the decays $\bd(t) \to D^{*\pm} \rho^\mp$,
$D^{*\pm} a_1^{\mp}$, $\dbarp~~\kbarp ~~$, etc.  Similarly, $\sin^2
\phi_3$ can be cleanly extracted from $\bs(t) \to D_s^{*\pm}
K^{*\mp}$, or simply performing an angular analysis of the decay mode
$B^\pm \to \dbarp ~~K^{*\pm}$ . Due to difficulties in measuring CP
phases with other methods, $\sin^2 (2\phi_1 + \phi_3)$ may well be the
second clean measurement, after $\sin 2\phi_1$,
to be measured at $B$-factories.

D.L. and R.S. would like to thank Prof. A.~I. Sanda and the local
organizers of BCP4 for financial support. We thank the organizers for
an exciting conference. The work of D.L. was financially supported by
NSERC of Canada.

\end{document}